\documentclass{aa}  

\usepackage[utf8]{inputenc}
\usepackage[T1]{fontenc}
\usepackage{graphicx}
\usepackage{txfonts}
\usepackage{xcolor}
\usepackage{comment}
\usepackage{array}
\usepackage{subcaption}
\usepackage{placeins}
\usepackage{afterpage}
\usepackage{placeins} 
\usepackage{float}
\usepackage{orcidlink}
\usepackage{soul}
\usepackage{ulem}

\usepackage{hyperref} % caricalo verso la fine del preambolo

\hypersetup{
  colorlinks=true,
  linkcolor=blue,
  citecolor=blue,
  urlcolor=blue
}
\newcommand{\source}{4U 1901+03}
\newcommand{\bat}{\textit{Swift}/BAT}

\newcommand{\nustar}{Nu\textit{STAR}}
\def \nbins {\ensuremath{N_{\textrm{bins}}}}

\newcommand{\afirst}{A$_{1}$\,}

\newcommand{\ecycl}{$E_{\mathrm{cyc}}$}

\newcommand{\pspin}{P$_{\mathrm{spin}}$\,}

\newcommand{\unilum}{erg s$^{-1}$\,}

\newcommand{\pfrms}{PF$_{\mathrm{rms}}$}

\newcommand{\uniflux}{{erg s$^{-1} \mathrm{cm}^{-2}$\,}}
\newcommand{\unibat}{{count s$^{-1} \mathrm{cm}^{-2}$\,}}
\newcommand{\pphighen}{PP$_{28-31}$}
\begin{document}

   \title{The elusive cyclotron line in 4U 1901+03: hidden, yet present}

   \subtitle{}

   \author{Elena Ambrosi\orcidlink{0000-0002-9731-8300}\inst{1}
          \and Antonino D'A\'i\orcidlink{0000-0002-5042-1036}\inst{1}
          \and Giancarlo Cusumano\inst{1}
          \and Carlo Ferrigno\inst{2}
          \and Ekaterina Sokolova-Lapa\orcidlink{0000-0001-7948-0470}\inst{3}
          \and Dimitrios K. Maniadakis\orcidlink{0009-0008-1148-2320}\inst{1,4}
          \and Antonio Tutone\orcidlink{0000-0002-2840-0001}\inst{1}
          \and Georgios Vasilopulous\orcidlink{0000-0003-3902-3915}\inst{5}
          \and Peter Kretschmar\orcidlink{0000-0001-9840-2048}\inst{6}
          \and Christian Malacaria\orcidlink{0000-0002-0380-0041}\inst{7}
          \and Fabio Pintore\orcidlink{0000-0002-3869-2925}\inst{1}
           }
\institute{INAF - IASF-Palermo, via Ugo La Malfa 153, 90146 Palermo, Italy  \email{elena.ambrosi@inaf.it} 
\and
Department of Astronomy, University of Geneva, Chemin d’Écogia 16, 1290 Versoix, Switzerland 
\and
Dr.\ Karl Remeis-Observatory and Erlangen Centre for Astroparticle Physics,
Friedrich-Alexander Universit\"at Erlangen-N\"urnberg,
Sternwartstr.\ 7, 96049 Bamberg, Germany
\and
Dipartimento di Fisica e Chimica Emilio Segrè, Università di Palermo, via Archirafi 36, 90123 Palermo, Italy
\and 
Department of Physics, National and Kapodistrian University of Athens, University Campus Zografos, GR$\sim$15784, Athens, Greece
\and
European Space Agency (ESA), European Space Astronomy Centre (ESAC),
Camino Bajo del Castillo s/n, 28692 Villanueva de la Ca\~{n}ada, Madrid, Spain
\and
INAF - Osservatorio Astronomico di Roma, Via Frascati 33, I-00078, Monte Porzio Catone (RM), Italy
}
   \date{}

% \abstract{}{}{}{}{} 
% 5 {} token are mandatory
 
  \abstract
{Cyclotron resonant scattering features in accreting X-ray pulsars are often difficult to detect, especially when shallow or variable. Recent studies have shown that combining spectral and timing analyses enhances their detectability.}
{We investigated the evolution of energy-resolved pulse profiles of the X-ray pulsar \source\ during its 2019 giant outburst, focusing on the 30–40 keV range where there have been disputed claims of a cyclotron line detection and its properties.}
{We analysed four \nustar\ observations of \source\, at different luminosities. We studied energy-resolved pulse profiles using harmonic decomposition, cross-correlation analysis, energy–phase maps, and pulsed-fraction spectra, and we used Bayesian spectral modelling to assess the presence and properties of a cyclotron line.}
{We detected significant spectral–timing variability in the 30–40 keV range, becoming stronger at lower luminosities. We found a pronounced drop in the pulsed fraction near $\approx35\,$keV only in the lowest accretion state and in the first harmonic of one intermediate-luminosity observation. Adopting a Bayesian informative approach on the spectral parameters, we find evidence for a cyclotron line in all the examined energy spectra, with an average centroid energy of $E_{\mathrm{cyc}} \approx 32$ keV, varying by only $\approx1.6\%$, with anti-correlation between line depth and luminosity.}
{We show that a spectral-timing combined approach is more sensitive than phase-averaged spectroscopy to shallow cyclotron features. The luminosity-dependent evolution of pulse profiles and cyclotron line depth point to a drastic change in the emission geometry and accretion flow structure.}

{}
\titlerunning{How the cyclotron line plays hide and seek}
\authorrunning{Ambrosi et al.}
   \maketitle
%
%-------------------------------------------------------------------
\section{Introduction}\label{sec:intro}
Be X-ray binaries (BeXRBs) represent the most numerous subclass of High-Mass X-ray Binaries, typically consisting of a neutron star in an eccentric orbit around a young, massive Be star. These systems are characterised by transient, high-luminosity X-ray outbursts, often triggered by the interaction of the compact object with the circumstellar decretion disk of the companion \citep[see e.g.][]{reig:2011}. During these events, the neutron star accretes material along magnetic field lines, forming accretion columns where the dynamics are governed by the interplay between the mass accretion rate ($\dot{M}$) and the strong magnetic field (B $\approx 10^{12-13} $G). The resulting emission pattern is fundamentally determined by the critical luminosity, $L_{\text{crit}}$ \citep{Basko1976, becker:2012, mushtukov:2015, mush_tsy_rev}: at $L < L_{\text{crit}}$, the flow is decelerated by Coulomb interactions (sub-critical regime), leading to a pencil-beam pattern dominated by emission escaping along the magnetic axis. Conversely, at $L > L_{\text{crit}}$ (super-critical regime), a radiation-dominated shock halts the infalling plasma, and photons diffuse through the column's lateral walls in a fan-beam dominated pattern.
The emerging spectrum, shaped by thermal and bulk Comptonization \citep{Becker2022}, is further modified by relativistic effects (gravitational redshift and light bending) and by the (possible) presence of Cyclotron Resonant Scattering Features (CRSFs). These absorption-like features are primary diagnostics of the magnetic field strength and the emitting region's geometry \citep{Staubert2019, Schwarm2017a}. However, the phase-averaged spectral analysis often suffers from the mixing of different spin-phase contributions, complicating the interpretation of the underlying beam pattern.
The emission pattern’s dependence on luminosity and energy results in complex, multi-peaked pulse profiles (PPs) that evolve non-trivially across different regimes \citep[see][for an overview]{Alonso-Hernandez2022}. This variability complicates phase-averaged spectral analysis, as it inevitably mixes different physical contributions. To decouple these effects, a combined timing and spectral approach might alleviate the spectral model degeneracy. As demonstrated by \cite{ppanda1} (hereafter FDA23), the Pulsed Fraction Spectrum (PFS), derived from high-resolution energy-resolved pulse profiles, can isolate local substructures associated with spectral features like the iron K$\alpha$ line and cyclotron resonant scattering features (CRSFs). This technique has successfully identified cyclotron line wings in V0332+53 \citep{ppanda2} and helped constrain the spin-phase modulation of the CRSF depth in 4U 1538$-$52 through physical modelling \citep{Sokolova-Lapa2023, ppanda3}\\
In this work, we applied and extended the method presented in FDA23 to analyse the 2019 giant outburst of \source\, a BeXRB with spin period \pspin\, $\approx \,2.7$ s \citep{2005ApJ...635.1217G}, discovered thanks to the \textit{Uhuru} satellite \citep{1976ApJ...206L..29F} during a giant outburst. Its orbital period is $\approx$\,22.58 d, with an almost circular orbit ($ e = 0.0144$) and a projected semi-major axis of a$_{x}\mathrm{sin}\ i = 104.236$ lt-s \citep{2020JHEAp..27...38T}. Since its discovery, it underwent four major outbursts, the last one in 2019, of which we analysed the \nustar\, data.
The structure of the paper is the following:
In Sec.~\ref{sec:analysis} we describe the data processing; extract the properties of the pulse profiles (Sec.~\ref{subsec:timing}), and their Pulse Fraction Spectra (Sec.~\ref{subsec:pfs}) and how they are connected the high energy spectra of each observation (Sec.~\ref{subsec:spectral}). 
Finally, the discussion is in Sec.~\ref{sec:discussion}, while summary and conclusions in Sec.~\ref{Sec:sum_and_concl}.
\section{Analysis}\label{sec:analysis}
\subsection{Data reduction and filtering}
\begin{figure}[!h]
    \centering
    \includegraphics[width=\columnwidth]{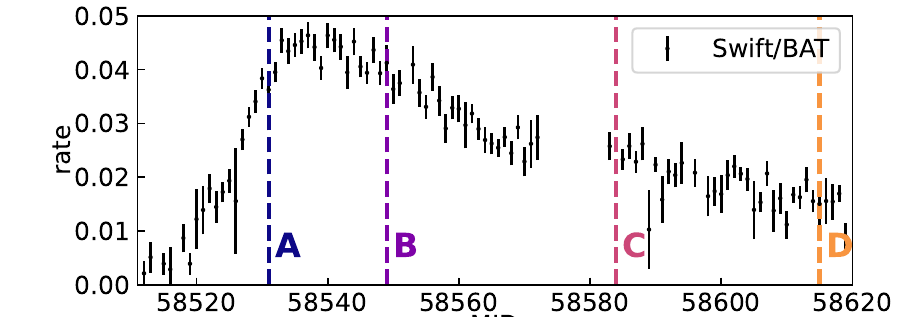}
    \caption{\bat\ light curve of the 2019 giant outburst (15-50 keV), rate in \unibat\ ; vertical lines mark the epochs of the four \nustar\, observations.}
    \label{fig:longterm}
\end{figure}
Our dataset consists of four \nustar\ observations, shown in Fig.\ref{fig:longterm} along with the 2019 outburst observed by \bat\ ($15 \-- 50\,$keV)\footnote{Swift/BAT transient monitor results
provided by the Swift/BAT team \citep{krimm:2013}}. One observation was performed before the peak of the outburst; the others were taken during the descending phase;
we will refer to these observations as Obs A to Obs D hereafter.
\begin{table}[ht] 
\caption{\nustar\ observations. Fluxes in the 4-50 keV
band are expressed in units of $10^{-9}$ \uniflux. \label{tab:log}}            % title of Table
     % is used to refer this table in the text
\centering                          % used for centering table
\begin{tabular}{l l l l}        % centred columns (4 columns)
\hline\hline    % inserts double horizontal lines
%\Tstrut\\
OBSID   & Expo$_{\mathrm{A/B}}$  & \multicolumn{1}{c}{\pspin}   & \multicolumn{1}{c}{Flux} \\
\small (label) &  \small (ks) &  \multicolumn{1}{c}{\small(s)} & \small  $10^{-9}\mathrm {erg}\ \mathrm{s}^{-1}\ \mathrm{cm}^{-2}$ \\
% table heading 

\hline                  % inserts single horizontal line
%\Bstrut \\

\tiny 90501305001 (A)   & \tiny17.8/18.4  &  \tiny  2.762529(5)  & \multicolumn{1}{c}{\tiny$\approx 6.9$}\\   
\tiny 90502307002 (B)  & \tiny12.3/12.5  &   \tiny 2.762251(6) & \multicolumn{1}{c}{\tiny$\approx 8.1$} \\
\tiny 90502307004 (C)   & \tiny21.5/21.6   &  \tiny 2.7618204(4) & \multicolumn{1}{c}{\tiny$\approx 4.5$ }\\
\tiny 90501324002 (D)   & \tiny45.1/45.4    & \tiny 2.7615742(2) & \multicolumn{1}{c}{\tiny$\approx 2.7$}\\
 
\hline                                   %inserts single line
\end{tabular}
\end{table}

We performed the data reduction following standardized procedures wrapped into the Python module \textsc{nustarpipeline}\footnote{\url{https://gitlab.astro.unige.ch/ferrigno/nustar-pipeline}} presented in FDA23. Here, we  outline  the main steps:\\
We first obtained calibrated level 2 event files of the Modules (FPMA and FPMB) and defined the source region as the one containing 95$\%$ of the source signal, centred on the best known coordinates of \source\ (RA, Dec (J2000)\,=\,19:03:39.39, 03:12:15.8); we defined background regions with the same size, located in a detector area free of contaminating sources. 

Then, to avoid the contamination from flares and/or dips, we analysed the light curve and discarded periods with high source variability, following the same procedure shown in FDA23. We obtained the spin frequency in each observation after baricentric and orbit corrections\footnote{Provided by the \textit{Fermi}-GBM pulsar database: \url{https://https://gammaray.nsstc.nasa.gov/gbm/science/pulsars.html} \cite{Malacaria2020}} of the event arrival times. We did not detect any spin derivative in any dataset and, therefore, we applied no corrective term.
The effective exposures and spin periods are reported in Tab.~\ref{tab:log}.
\subsection{Energy resolved pulse profiles and energy-phase maps}\label{subsec:timing}
We applied the methods described and presented in FDA23 to compute the energy-resolved pulse profiles, but for this analysis we increased the energy resolution by a factor of 3 at lower energies (E < 10 keV), where the signal-to-noise ratio (S/N) is higher with respect to higher energies, ending with a total of 131 energy bins in the 2.0--70 keV full band.\\
\begin{figure*}
    \includegraphics[width=0.5\textwidth]{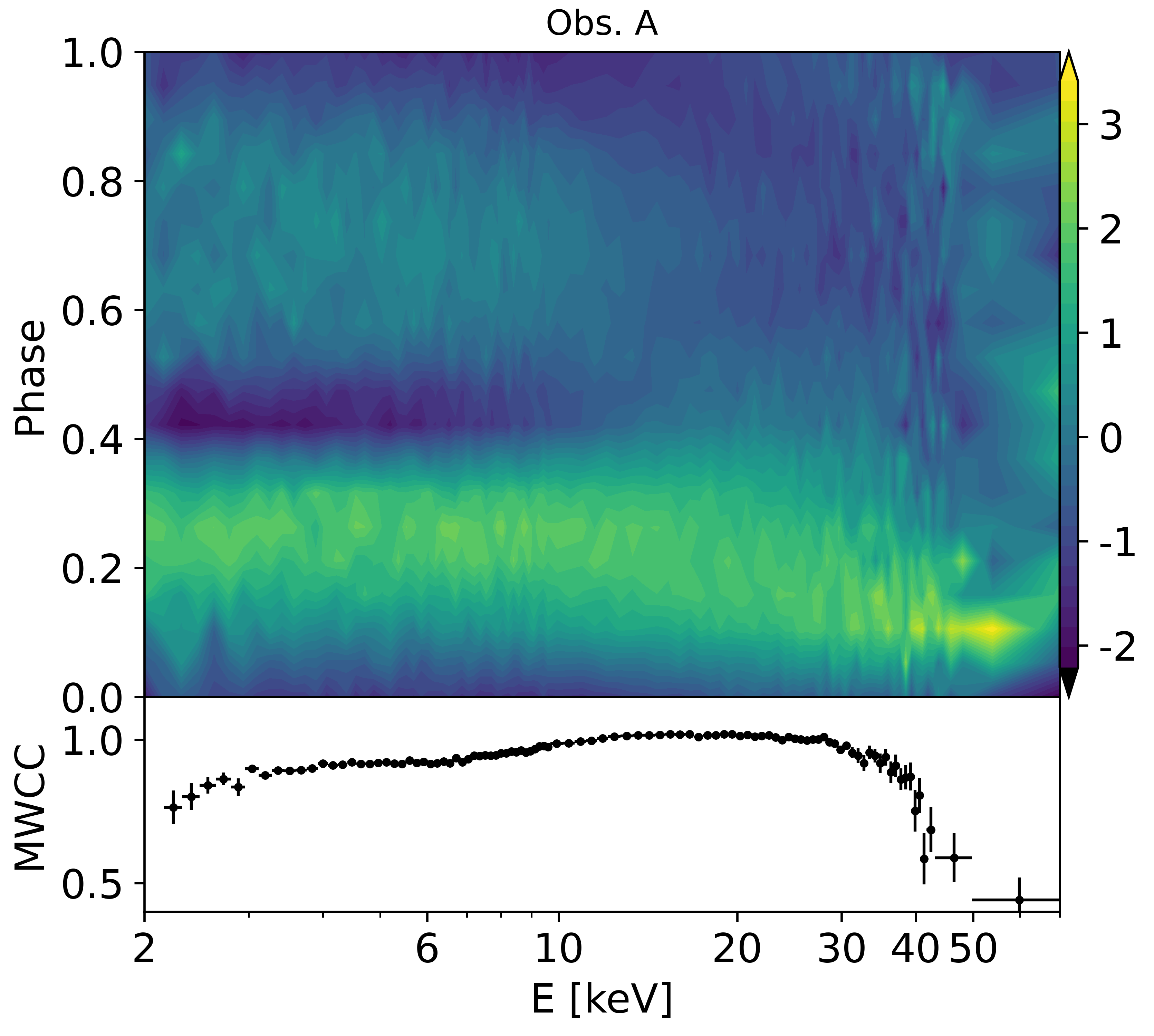}
    \includegraphics[width=0.5\textwidth]{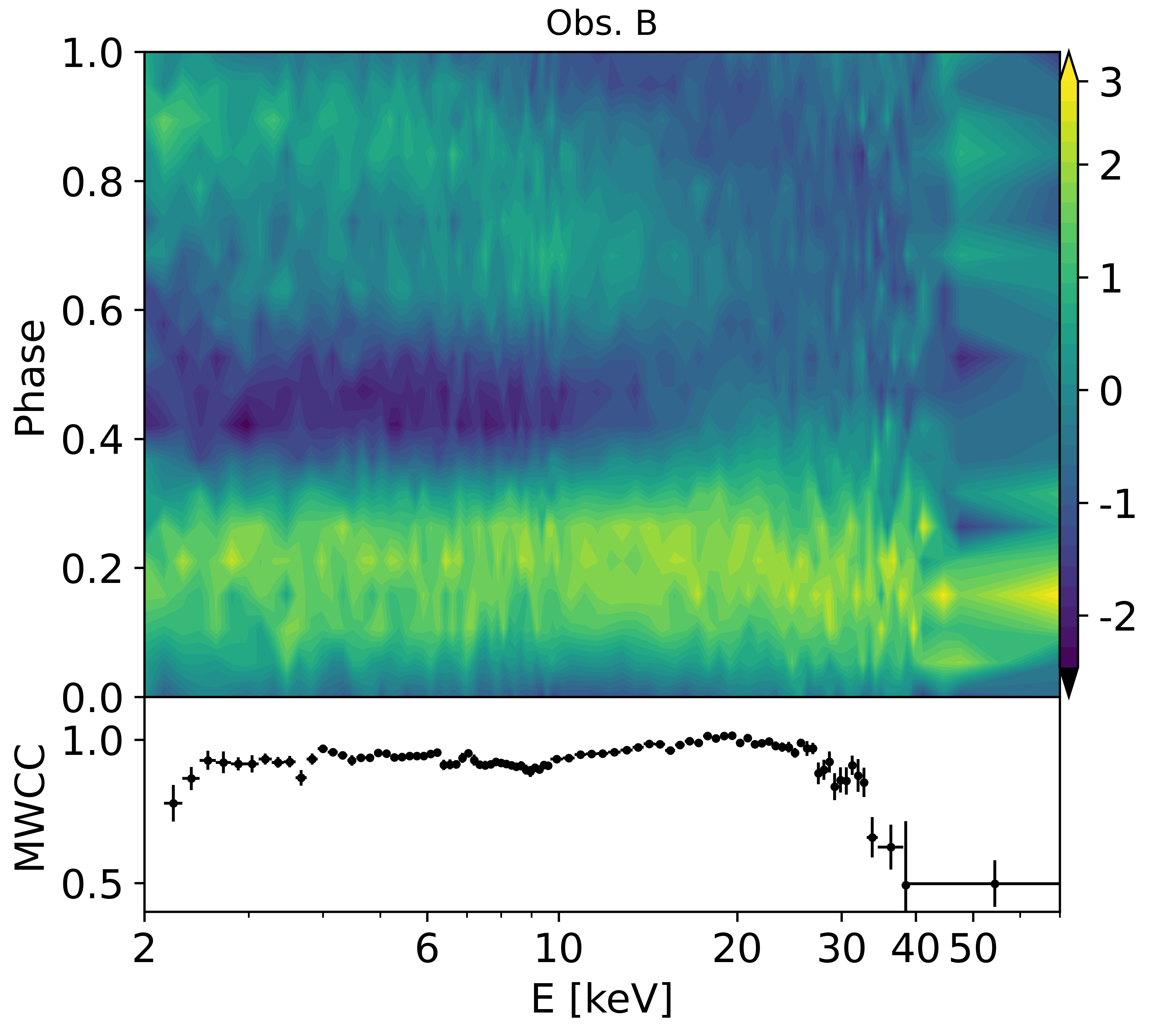}
    \includegraphics[width=0.5\textwidth]{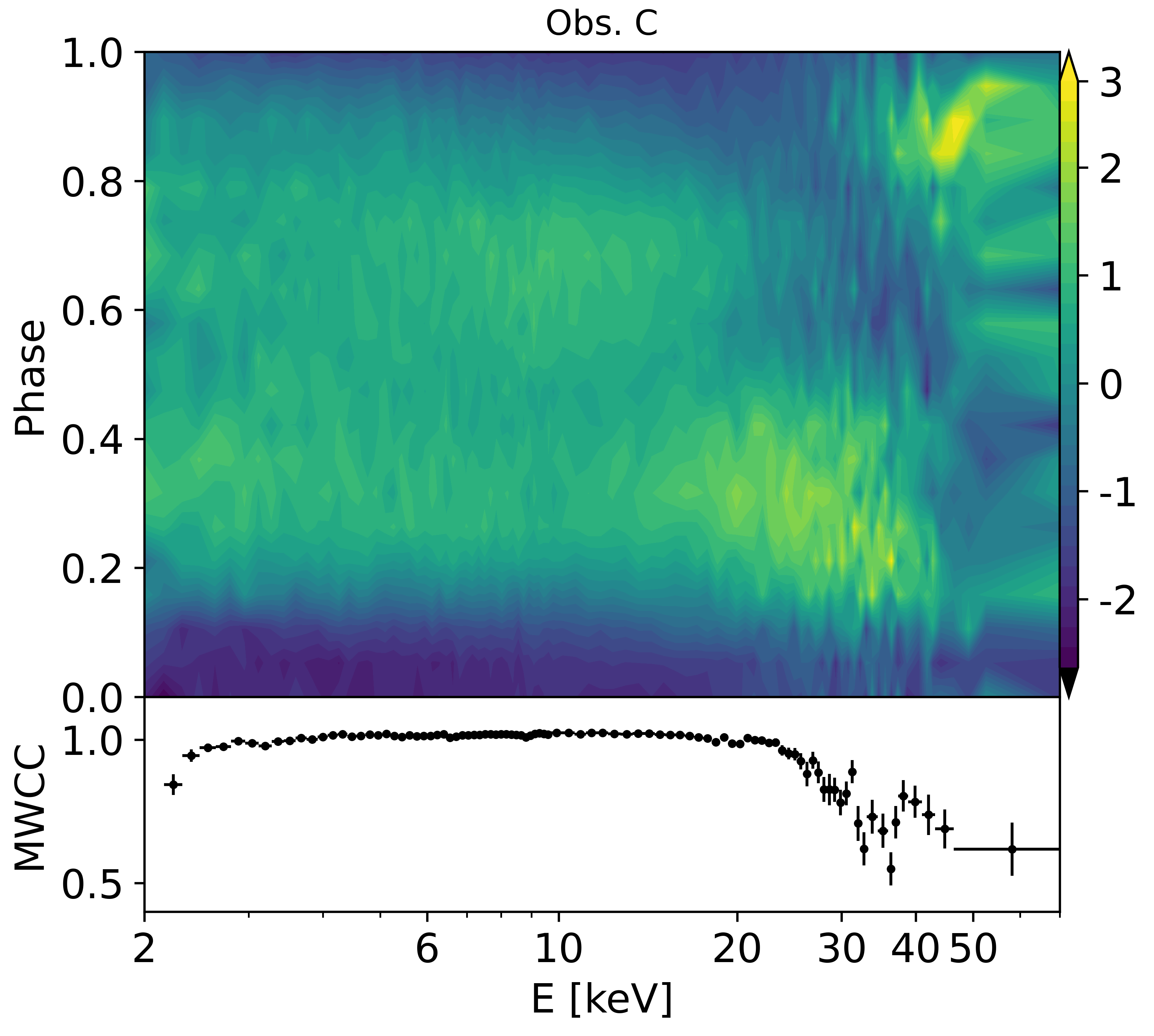}
    \includegraphics[width=0.5\textwidth]{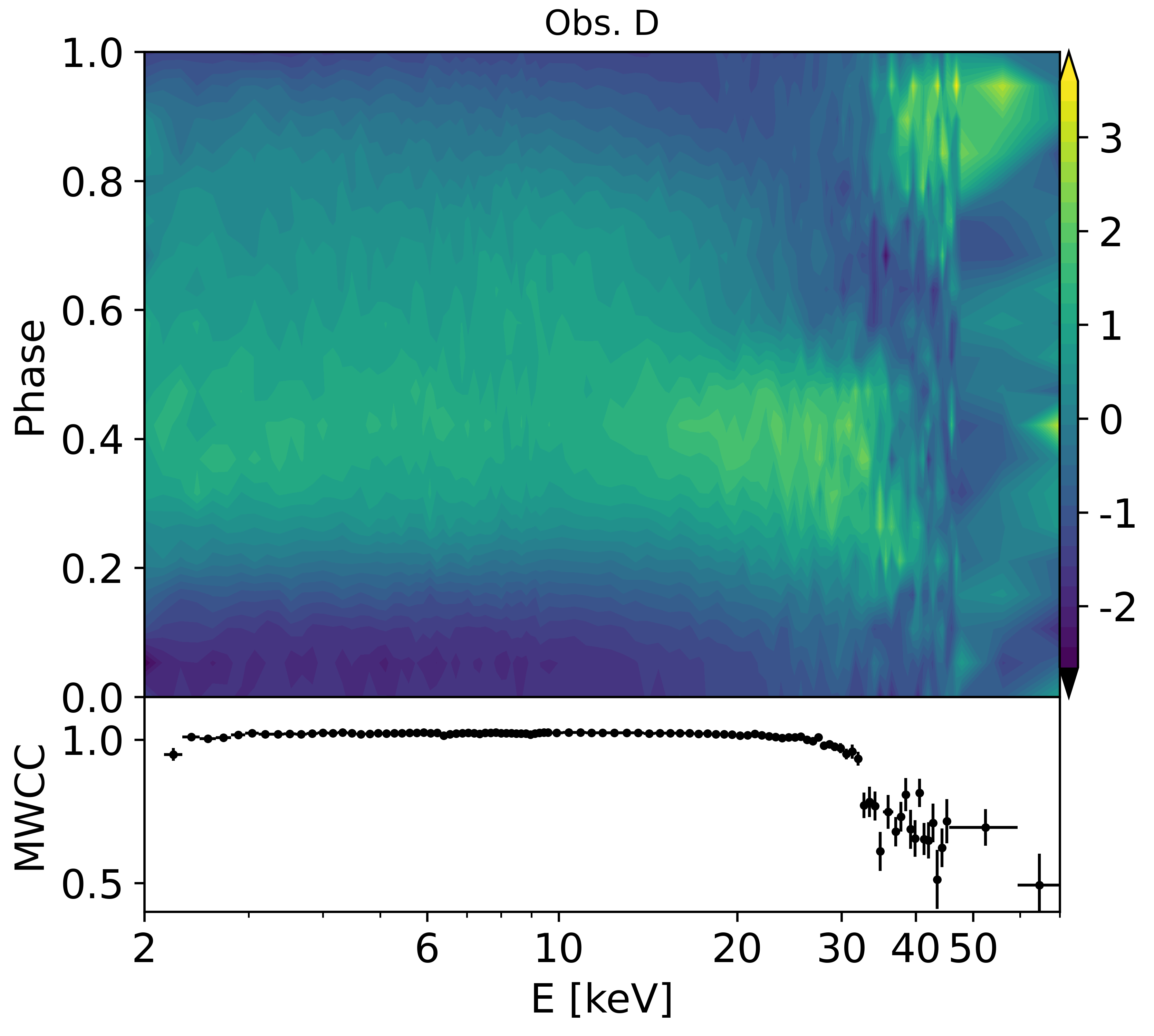}
    \caption{Energy-phase maps and correspondent moving-window cross-correlation (MMWC) with a window of 2 adjacent bins (see text). For each panel, we chose \nbins = 20 and S/N = 5.0. The color bar, on the right side of each map, represents the values of the normalised profiles plotted in the map and spans between the minimum and the maximum of the entire map.}% Phase 0 corresponds to the minimum of the 3-70 keV pulse profile, like the energy-averaged PPs in Fig.\ref{fig:broad_band_profiles}.}
    \label{fig:epmaps_1}
\end{figure*}
We extracted energy-resolved pulse profiles with 20 phase bins and re-binned them in energy to reach a minimum S/N of 5. A good visual representation is given through the Energy Phase Maps (hereafter EPM), which illustrate at once the pulse shape variations with energy and phase, as shown in Fig.~\ref{fig:epmaps_1}. We show them in the upper panels of Fig. \ref{fig:epmaps_1} with the phase in the y-axis and energy in the x-axis  (pulse profiles extracted in some representitave energy bands for the four dataset are shown in \citealt{Beri2021}).
To better investigate the local variability of the energy-resolved pulse profiles, we employed a moving window cross-correlation (MWCC) technique, designed to track rapid changes between adjacent profiles. We adopted the same method described in FDA23 (see their Eq. 3 and Fig. 3), with the difference that here each energy-resolved profile is compared with a reference profile obtained by averaging the surrounding profiles, excluding the energy bin of interest, within a maximum index distance of $w$.
After some testing, we found that $w$\,=\,2 provides the most suitable balance between sensitivity to local variations and statistical robustness, allowing us to effectively trace profile changes across the entire energy range.
$\mathrm{MWCC}$ values  are shown in the bottom panels of each plot in Fig. \ref{fig:epmaps_1}.
The most important result emerging from these four plots is that they share a common behaviour above 
30 keV, where the  MWCC values exhibit an abrupt drop. Below this energy, Observations C and D show a high degree of similarity, with MWCC values approximately close to unity, whereas the higher-luminosity observations (A and B) display significant energy-dependent variability in their pulse profiles.

\subsection{Pulsed Fraction Spectra}\label{subsec:pfs}
As shown in FDA23, the parallelism between the energy spectra and the pulsed fraction spectra can be quantified by modelling the latter. We defined a 'continuum', described by a polynomial, to trace the general trend of the pulsed fraction spectrum, while local substructures at the energies of the iron K$\alpha$ line (6.4\,keV) and the CRSF were modelled with negative Gaussians, which we referred to as ‘dips’.

\begin{figure*}[]
    \centering
    \includegraphics[width=\columnwidth]{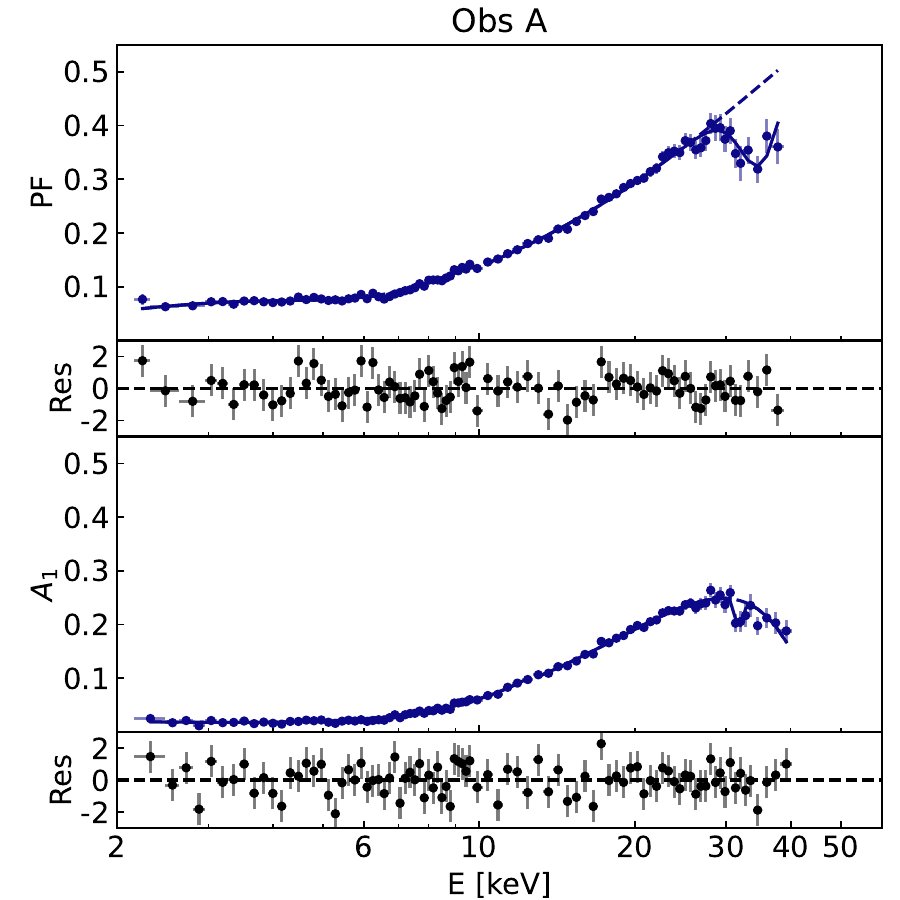}
    \includegraphics[width=\columnwidth]{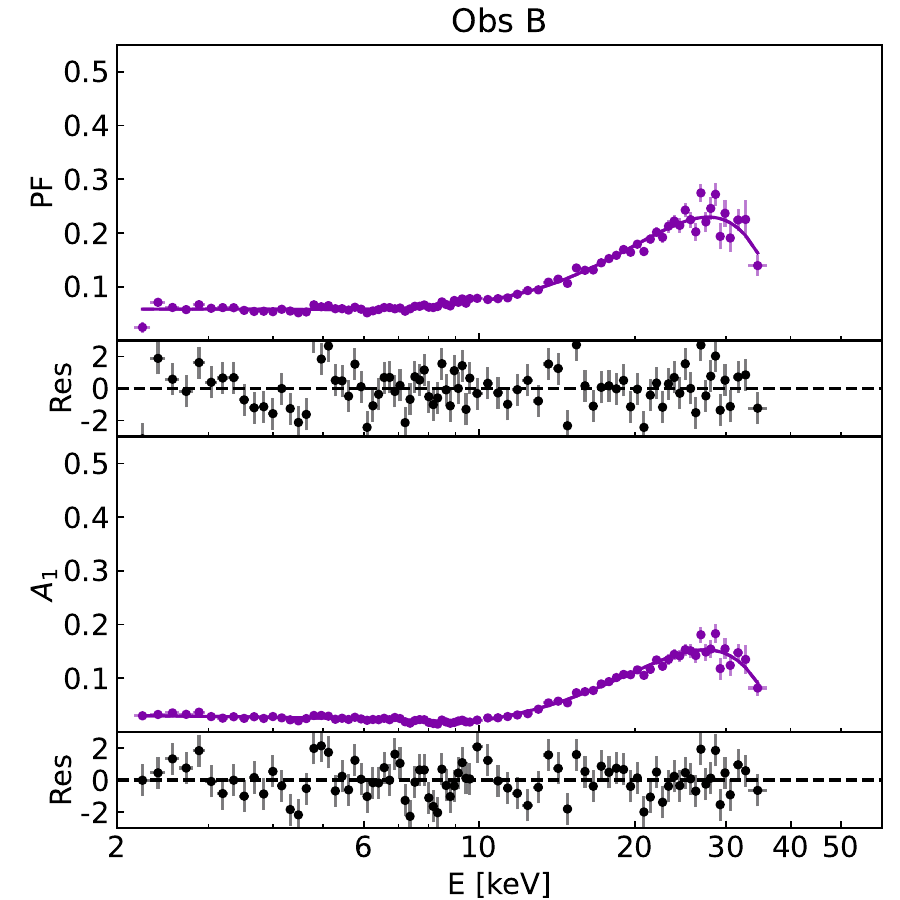}
    \includegraphics[width=\columnwidth]{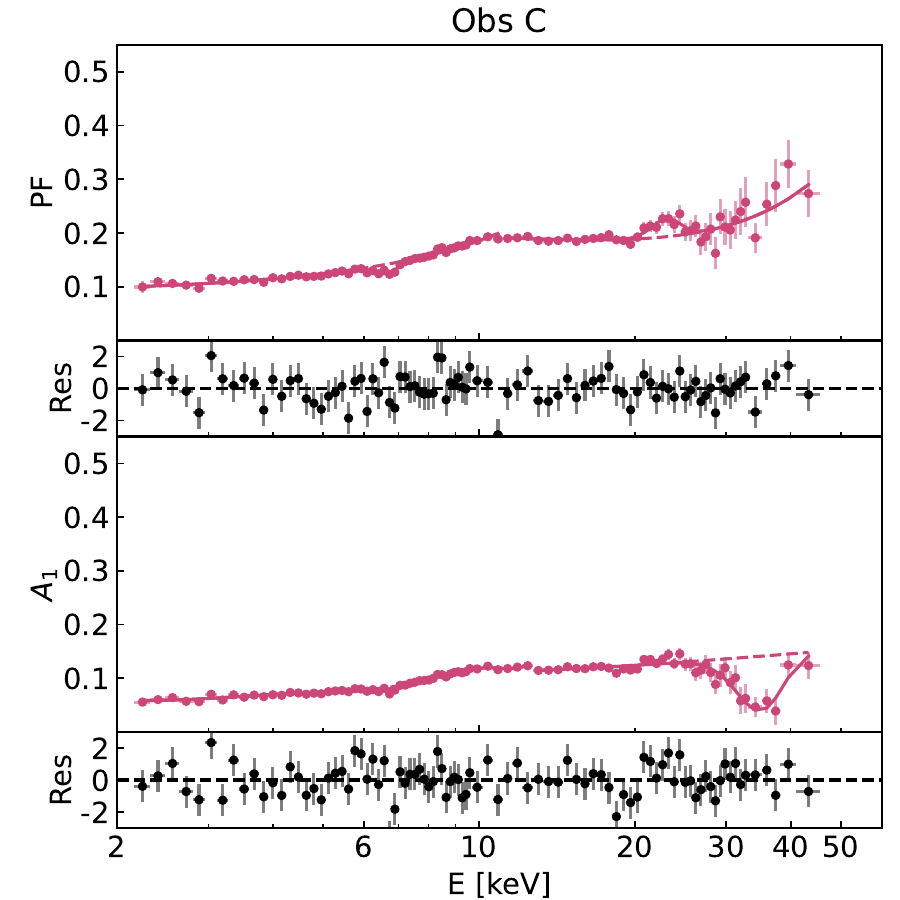}
    \includegraphics[width=\columnwidth]{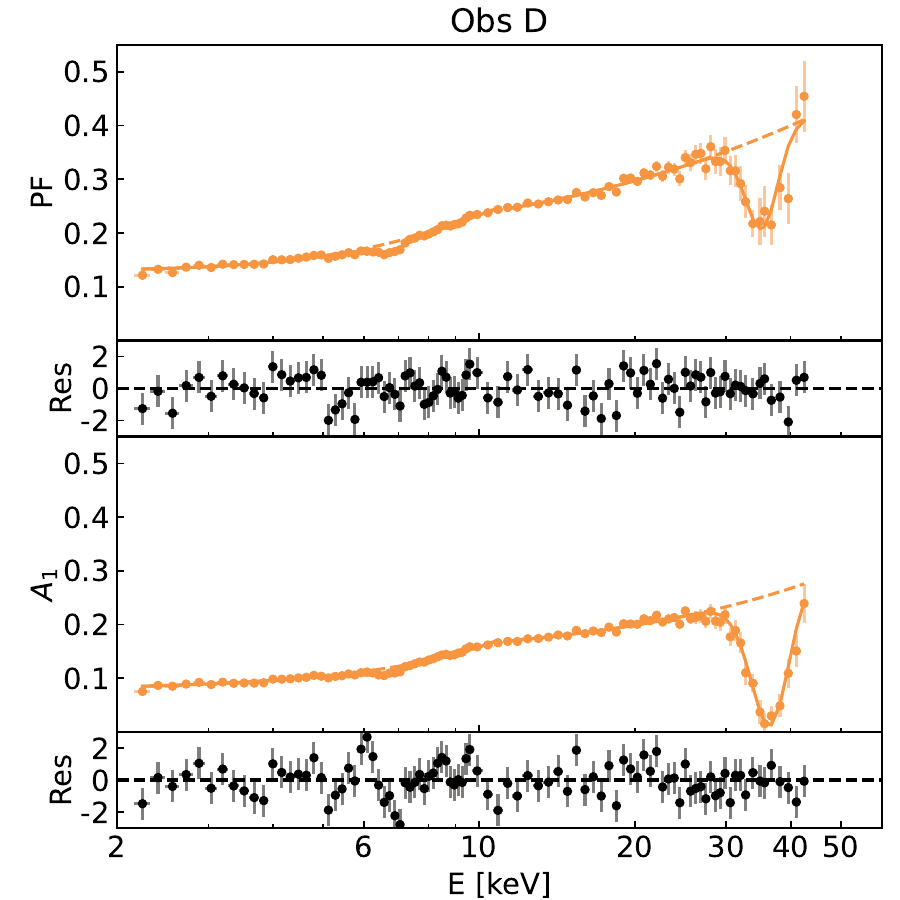}
    \caption{PFS and first harmonic amplitude (\afirst) (data points) for Obs A-D: dashed lines represent the best fit continuum, solid line the best fit model (continuum + gaussian in absorption/emission) and residuals (black points).}
    \label{fig:fit_pfAndharms}
\end{figure*}

We computed the PFS of \source\ using the root-mean-squared (rms) definition\footnote{See FDA for an extensive analysis of the different definition of pulsed fraction.} of the pulsed fraction, \pfrms\
which quantifies the variability of each profile relative to its mean. In practice: 
\begin{equation}\label{eq:pf_rms_c}
PF_{\mathrm{rms}} = \frac{\sqrt{ \Sigma_{i=0}^N \left[ (c_i - \bar c)^2 - \sigma_{c_i}^2 \right] / N}}{\bar c}
,\end{equation}
Here, $\bar c$ is the average count rate for each profile, $c_i$ is the normalised count rate in the $i$ phase bins and $\sigma_{c_i}$ the corresponding standard deviation. 
Beyond the global variability traced by the PFS, the harmonic components of the pulse profiles can provide additional insights. In particular, we analysed the energy dependence of the amplitude of the first harmonic, obtained from the Fourier decomposition of each profile, and modelled it in the same way as for the PFS.\\
In Fig.~\ref{fig:fit_pfAndharms} we show the PFS and the spectrum of the normalised amplitude of the first harmonic for all observations.
We find strong similarities between Obs A and Obs B, and likewise between Obs C and Obs D, indicating that the source behaviour groups according to luminosity. This pattern mirrors the luminosity-dependent trend already observed in the energy–phase maps (Fig.~\ref{fig:epmaps_1}).
\subsubsection{Phenomenological fit of the pulsed fraction spectra}
We performed a phenomenological fit of the PFS of the four observations searching for hints of local features in the iron line region ($6\--7$\,keV) and in the $30\--40$\,keV range, where the energy-phase matrix showed sharp variability among the profiles, indicating the possible presence of a CRSF. In addition to the pulsed fraction spectra, we extended our analysis to include the normalised amplitude of the first harmonic, \afirst, in line with the findings of FDA23.\\
Inspection of the four PFSs shown in Fig. 3 reveals that Obs D exhibits a clear dip in the $30\--40$\,keV range. In contrast, this complexity is much less pronounced in the other three observations, for which we nevertheless conducted a dedicated search for similar features. All fits required dividing the data into two distinct energy regions, with E$_{\rm split}$ lying in the $9\--11$\,keV range (see FDA23 for details), hereafter referred to as the Low-Energy and High-Energy intervals.
For each interval, we tested both a simple model, consisting of a polynomial continuum, and a more complex model including an additional Gaussian component. We emphasize that the polynomial is used purely for descriptive purposes, without assuming any specific physical model. Its flexibility provides an empirical representation of the PFS continuum, allowing the identification of local deviations potentially associated with spectral features.
To avoid overfitting, we selected the best fit model as the one which minimizes the Bayesian Information Criterion (BIC, \citealt{Schwarz1978}) that, under the assumption of Gaussian errors, can be written as:
\begin{equation}
    \mathrm{BIC} =  \chi^{2}_{min} + k\ln{n}
\end{equation}
where \textit{k,n} are the number of free parameters in the model and the data, respectively. Table~\ref{tab:PFSfit} shows the differences in BIC values as $\Delta\ \mathrm{(BIC)} = \mathrm{BIC}_{\mathrm{worst}} - \mathrm{BIC}_{\mathrm{best}}$.
 \begin{table*}[htbp]
\centering
\caption{Best-fit parameters of the Gaussian feature detected at high energies for the PFS and its first harmonic (A$_1$).
%Reported are the medians with 68\% credible intervals (99.7\% in square brackets) and the model significance of the model with the gaussian with respect to simple polynomial $\Delta$(BIC); Obs B is omitted as its PFS shows no significant features.
}
\label{tab:PFSfit}
\resizebox{0.98\textwidth}{!}{
\begin{tabular}{l| l l | l l | l l}
\hline
\textbf{High Energy} &   &  &   &    &   & \\ 
 & \multicolumn{2}{c|}{Obs. A}  & \multicolumn{2}{c|}{Obs. C}  & \multicolumn{2}{c}{Obs. D} \\
 %\vspace{0.1cm}
 &   \multicolumn{1}{l}{PFS} & \multicolumn{1}{l|}{\afirst} & \multicolumn{1}{l}{PFS} & \multicolumn{1}{l|}{\afirst} &  \multicolumn{1}{l}{PFS} & \multicolumn{1}{l}{\afirst} \\

 $E$ (keV) 
& $34.9 \pm 0.5 \; [-1.6, +1.9]$ 
&  $37.9_{-5.4}^{+2.3} $ 
& $23.1 \pm 0.5 \, [-1.9, +5.2]$
& $34.9 \pm 0.6 \; [\pm 2.0]$
& $35.22 \pm 0.34 \; [\pm 1.0]$ 
& $36.46 \pm 0.19 \; [\pm 0.56]$ \\
$\sigma$ (keV) 
& $2.8^{+0.1}_{-0.3} \; [-1.14, +0.19]$ 
&  $2.3_{-0.8}^{+0.5} $
& $1.7_{-0.4}^{+0.5}\, [-1.6, +0.9]$
& $3.5^{+0.3}_{-0.5} \; [-1.6, +0.5]$ 
& $1.94^{+0.04}_{-0.09} \; [-0.40, +0.06]$ 
& $2.95^{+0.04}_{-0.08} \; [-0.33, +0.05]$ \\
Amplitude 
& $-0.9 \pm 0.2 \; [\pm 0.6]$ 
& $0.1_{-0.3}^{+0.7}$
&  $0.13_{-0.04}^{+0.05}\, [-0.24, +0.15]$
&$-0.8 \pm 0.2 \; [-0.6, +0.5]$
& $-0.9 \pm 0.1 \; [\pm 0.3]$ 
& $-1.80 \pm 0.09 \; [-0.20, +0.27]$ \\
\hline
\multicolumn{1}{l|}{$\Delta$ (BIC)}   & \multicolumn{1}{l}{$ 23.3 $} & \multicolumn{1}{l}{$1.4$ } &  \multicolumn{1}{l}{$ 34.6 $} & \multicolumn{1}{l|}{$21.7$} & \multicolumn{1}{l}{$ 28.4 $} & \multicolumn{1}{l}{$  61.0 $ }
 \end{tabular}
 }
 \tablefoot{Reported are the medians with 68\% credible intervals (99.7\% in square brackets) and the model significance of the model with the gaussian with respect to simple polynomial $\Delta$(BIC); Obs B is omitted as its PFS shows no significant features.}
\end{table*}   
Synthesizing the model results, we find that:
\begin{itemize}
\item $\mathrm{PFS_A}$ requires a complex model, significantly improving the fit ($\Delta \mathrm{BIC} = -23.3$), with a Gaussian at $E_{\mathrm{gau}} \approx 34.9 \mathrm{keV}$ ($\sigma \approx 2.8$\,keV). 
\item For $\mathrm{PFS_B}$, limited to $E \lesssim 35$\,keV, the data are simply well fitted with a polynomial.
\item $\mathrm{PFS_C}$ need a negative Gaussian at $6.6$\,keV; at higher energies, residuals are mitigated by a positive Gaussian at $\sim 23$\,keV, although its parameters are unconstrained. The $\mathrm{PFS_C}$ first harmonic instead requires a negative Gaussian at $\sim 34$\,keV.
\item The last observation ($\mathrm{PFS_D}$) shows the most significant features: both $\mathrm{PFS_D}$ and its first harmonic increase with energy but exhibit sharp Gaussian-like drops at low energy ($E_{\mathrm{Fe}} = 6.67 \pm 0.06 ,\mathrm{keV}$, $\sigma_{\mathrm{Fe}} = 0.53 \pm 0.14$) and at high energy ($E_{\mathrm{gau}} = 35.2 \pm 0.3$\,keV, $\sigma = 1.94^{+0.04}_{-0.09}$, depth $\approx -1.2$, see Tab.~\ref{tab:PFSfit}).
\end{itemize}
\subsection{Spectral analysis}\label{subsec:spectral}
\subsubsection{Methods}
\begin{figure*}
\includegraphics{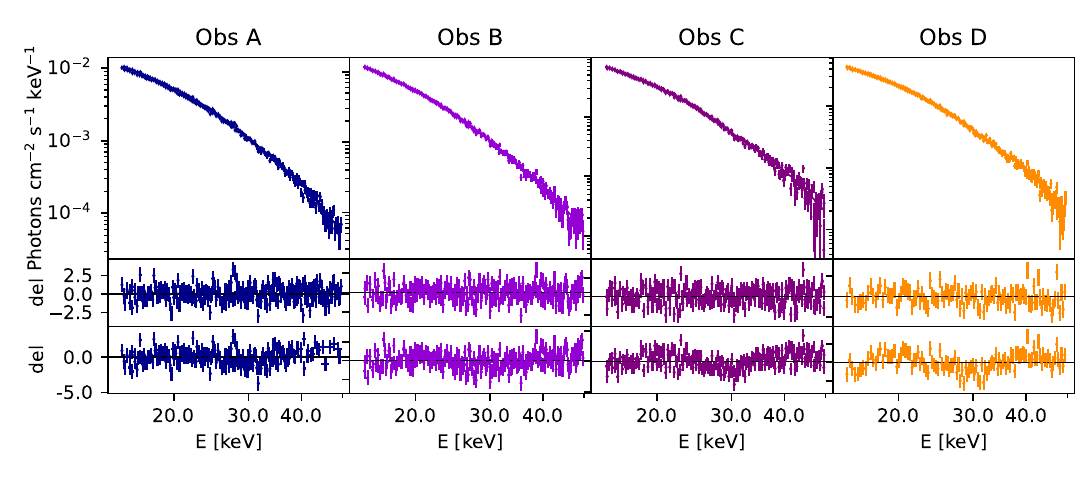}
\caption{(Upper panels): data and best-fitting model (Model2) for the phase averaged-spectra of the four \nustar\ observations; residuals of best-fitting model (middle panels); residuals of the best fitting model with a line depth set to zero (lower panels) }\label{fig:phase_ave}
\end{figure*}
In the following we test the hypothesis that the sharp variability observed near 35 keV, evident in the pulse profiles (Fig.~\ref{fig:epmaps_1}), Pulsed Fraction Spectra, and the first harmonic's normalised amplitude (Fig.~\ref{fig:fit_pfAndharms}), originates from a cyclotron resonance. We performed the spectral analysis in the restricted energy range $15\--50$\,keV, where the continuum can be well approximated by an exponentially cut-off power-law ($\displaystyle A(E) = K E^{-\Gamma} \exp(-E/E_{\mathrm{cut}})$, \textsc{cutoffpl} in \textit{Xspec}), thus minimising the spectral correlations and difficulties involved by the choice of a more complicated  broad-band continuum.%, which would require many additional parameters.
We used the widely adopted X-ray fitting package \textit{Xspec} \citep{arnaud1996xspec}, interfaced via its \textit{Python} wrapper \textit{PyXspec}. 
We compared the fit results of this simple model with those obtained with the same model convolved with a Gaussian absorption feature:\\
i) Model1 = $\textsc{const} \times (\textsc{cutoffpl})$; \\
ii) Model2 = $\textsc{const} \times (\textsc{cutoffpl} \ast \textsc{gabs})$. \\
To go beyond the standard frequentist statistic we employed a Bayesian statistical framework to perform model selection and rigorously assess model uncertainties. 
For this purpose, we used the BXA package (Bayesian X-ray Analysis; \citealt{Buchner:2014}), which connects \textit{Xspec} with the \textit{UltraNest} nested sampling algorithm \citep{Buchner:2021}, allowing efficient computation of posterior distributions. Moreover, it computes the marginal evidence for each model, that is, the likelihood integrated over the prior space:
\begin{equation}
Z = \int \mathcal{L}(\theta)\,\pi(\theta)\, d\theta,
\end{equation}
where $\mathcal{L}(\theta)\,$ and $\pi(\theta)\,$ are the likelihood and the prior distribution, respectively.
We then identified the preferred model as the one with the highest marginal evidence and quantified the relative support provided by the data through the Bayes Factor:
\begin{equation}
 \mathrm{BF} = \left( \frac{Z_{\mathrm{best}}}{Z_{\mathrm{model}}} \right)
\end{equation}
where $Z_{\mathrm{best}}$ is the evidence of the best-performing model, while  Z$_{\mathrm{model}}$ is the evidence of the other model. Expressing the Bayes Factor in logarithmic form, as $\log_{10}(\mathrm{BF})$, we adopted the Jeffreys scale \citep{jeffreys_bf} to interpret its magnitude, considering $\log_{10}(\mathrm{BF}) \geq 1$ as strong evidence and $\log_{10}(\mathrm{BF})\geq 2$ as decisive evidence in favour of the model with the larger marginal evidence.
\subsubsection{Results}
We rebinned the spectra using a \textit{Python} implementation\footnote{see \url{https://gitlab.astro.unige.ch/ferrigno/optimal-binning}} algorithm for optimal binning, following the prescriptions of \cite{optim_bin_kaastra}.
We adopted uninformative yet physically motivated priors for the parameters of the \textsc{gabs} component, applying a single set of broad priors, motivated by the timing analysis results, across all observations. Specifically, the prior intervals for \textsc{gabs} were: \ecycl\ = [28.0, 38.0] keV, $\sigma$ = [0.1, 7.0] keV, and Depth = [0.01, 7.0], as reported in Table~\ref{tab:priors_phase_av} along with their corresponding probability distributions. We assumed uniform distributions for all parameters, except for the \textsc{cutoffpl} normalization and the depth of the Gaussian, for which we adopted log-uniform distributions.
\begin{table}
\small
\caption{Set of priors adopted for the phase-averaged spectral analysis with both Model 1 and Model 2.}
\label{tab:priors_phase_av}
    \begin{tabular}{l|l|l}
    \hline
     Parameter & [min, max, start, step]& Prior distribution \\
     \hline
      C$_{FPMB}$  & [0.8, 1.3, 0.99, 0.01]  &   Uniform \\
      $\Gamma_{\mathrm{cutoffpl}}$  & [-4.0, 6.0, -0.25, 0.1]  &  Uniform\\
      $E_{cut, \mathrm{cutoffpl}}$  & [1, 10, 6.4,  0.1]  & Uniform \\
      Norm$_{\mathrm{cutoffpl}}$  &  [$1\cdot 10^{-10}$, 1.,$5\cdot 10^{-2}$, 0.01] &  Log-uniform  \\
      \ecycl &  [28.0, 38.0, 31.8,  0.1]  &  Uniform \\ 
      $\sigma$ &  [0.1, 7.0, 3.9,  0.1]  &   Uniform\\
      D & [0.01, 7.0, 0.7, 0.01] & Log-uniform \\
      \hline
    \end{tabular}
\end{table}
%The set of the prior distribution for each parameter is shown in Tab~\ref{tab:priors_phase_av}.\\
\begin{table}[]
\caption{Posterior parameter estimates for the cyclotron resonance model.}
%\centering
\resizebox{\columnwidth}{!}{%
    \begin{tabular}{l  l  l  l l} 
    \hline
    \vspace{0.03cm}
       Parameter & A & B  & C & D    \\
       \hline
       \vspace{0.03cm}
       $C_{FPMB}$ & $0.938 \pm 0.003$ & $0.988 \pm 0.004$ & $0.985 \pm 0.004$ & $0.969 \pm 0.003$  \\
       \vspace{0.03cm}
      $\Gamma$ & $-0.27 _{-0.08}^{+0.07}$ &  $0.2 \pm 0.1$ & $-0.1\pm 0.1$ & $-0.8\pm 0.1$ \\
      \vspace{0.03cm}
      E$_{cut}$\,[keV] & $6.2 \pm 0.1$ & $6.7 \pm{0.2} $ & $6.1 \pm 0.2$ & $5.4 \pm{0.1}$  \\
      \vspace{0.03cm}
      $\mathrm{Norm}$ & $0.057 \pm 0.009$ & $0.20 \pm 0.06$ & $0.06 \pm 0.01 $  & $0.007 \pm 0.002$\\
      \vspace{0.03cm}
     \ecycl\,[keV] & $32.1^{+1.2}_{-0.8}$ & $32.2 \pm 1.3$ & $32.1 \pm 0.8$ & $31.7\pm 0.6 $\\
     \vspace{0.03cm}
     $\sigma_{CRSF}$\,[keV] &  $4.5 _{-1.3}^{+1.4}$ & $6.0^{+0.7}_{-1.0}$ &  $5.2 _{-0.8}^{+0.9}$ & $5.8 \pm 0.7$ \\
     \vspace{0.03cm}
     $\mathrm{Depth}_{\mathrm{CRSF}}$\,[keV] &  $0.8^{+0.5}_{-0.3}$ & $1.6^{+0.6}_{-0.5}$ & $3.0^{+1.1}_{-0.8}$ & $2.7_{-0.6}^{+0.8}$  \\ 
     \hline
    log$_{10}\mathrm{BF}$   &  0.98 & 5.1  & 25.0  & 23.8\\ 
\end{tabular}
}
\tablefoot{We report the median and 68\%  credible intervals (16th and 84th percentiles) for each parameter. The final row provides the log-evidence ($\log_{10}\text{BF}$) compared to a null model.}
\label{tab:bayesian_fit}
\end{table}
For all observations, the model that provides the best description includes the \textsc{gabs} component, even if for obs A the simple model is only one order of magnitude less probable. The Bayes factors are log$_{10}\mathrm{BF}$ = 0.98, 5.1, 25.0, 23.1 for observations A, B, C and D, respectively, indicating decisive support for the presence of the line in observations B, C, and D and indication for it in observation A. 
Fig.~\ref{fig:phase_ave} shows the phase-averaged spectra and the residuals for the best fit. For comparison, we also show the residual after setting  the depth of the line to zero.
\subsection{Pulse Profiles variability at \ecycl}\label{sec:pp30_40}
\begin{figure*}[t]
    \captionsetup{font=small, skip=3pt}
    \centering
     \includegraphics[height=0.3\textheight]{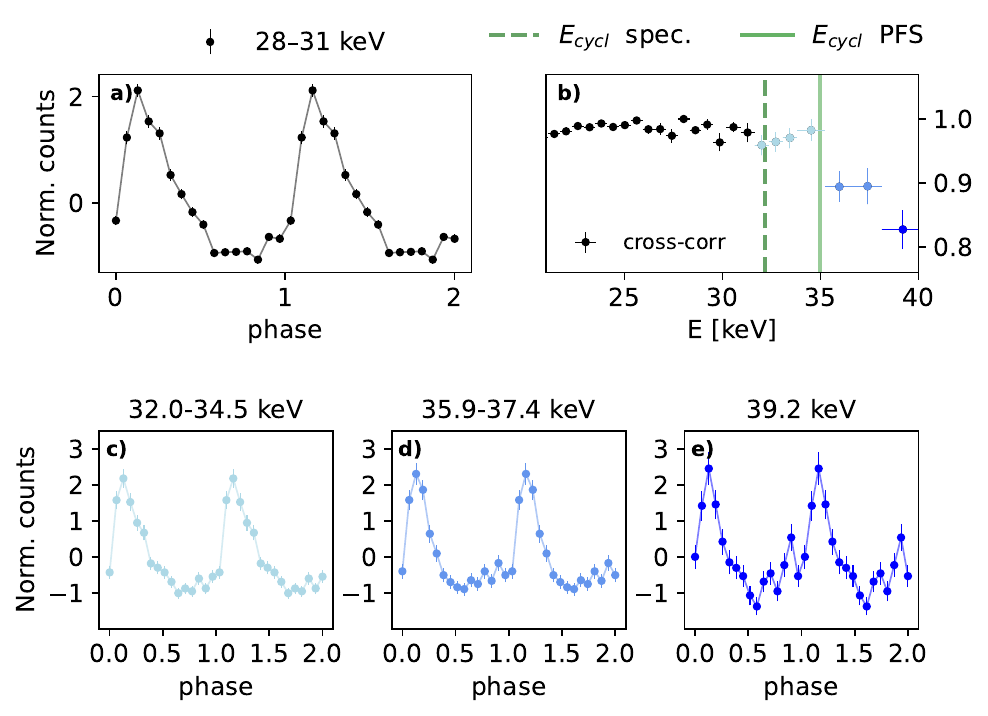}
    \vspace{2mm}
    \includegraphics[height=0.3\textheight]{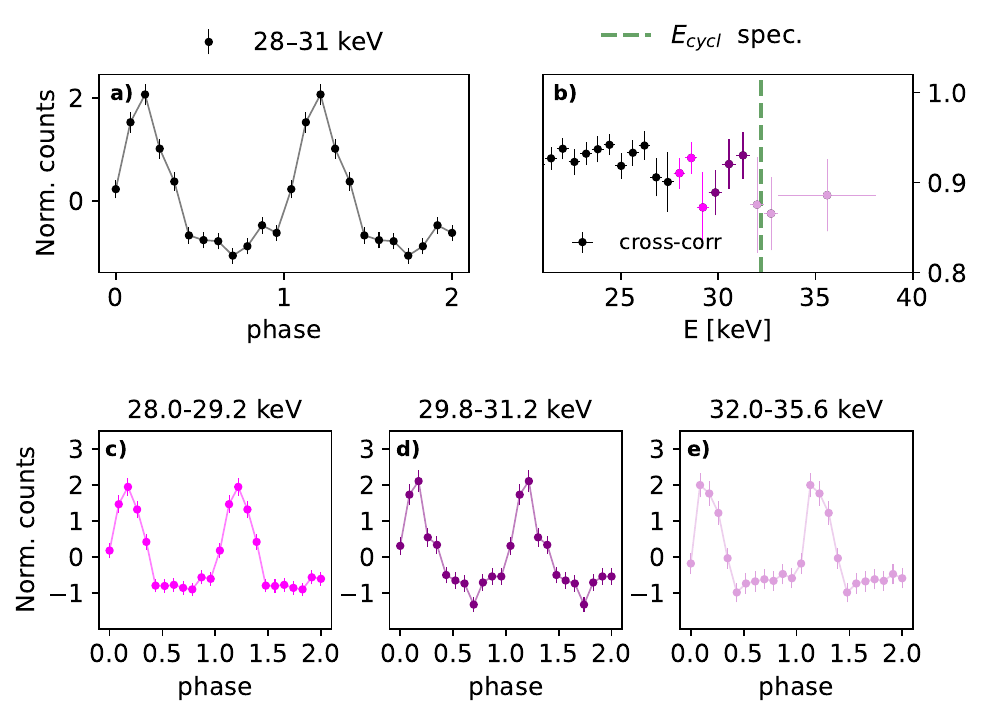}
    \caption{Pulse variability around the cyclotron line for observations A (top, blue tones) and B (bottom, purple tones). a)\pphighen (see Sec.~\ref{sec:pp30_40}); b) cross-correlation spectra of the PPs > 20 keV relative to \pphighen, dashed and solid vertical lines: cyclotron energy from the spectral and PFS fits, respectively. c),d),e): PPs in three energy ranges with consistent cross-correlation values, using the same color code adopted in b).}
    \label{fig:pps}
\end{figure*}
\begin{figure*}[h!]
\centering
\includegraphics[height = 0.43\textheight]{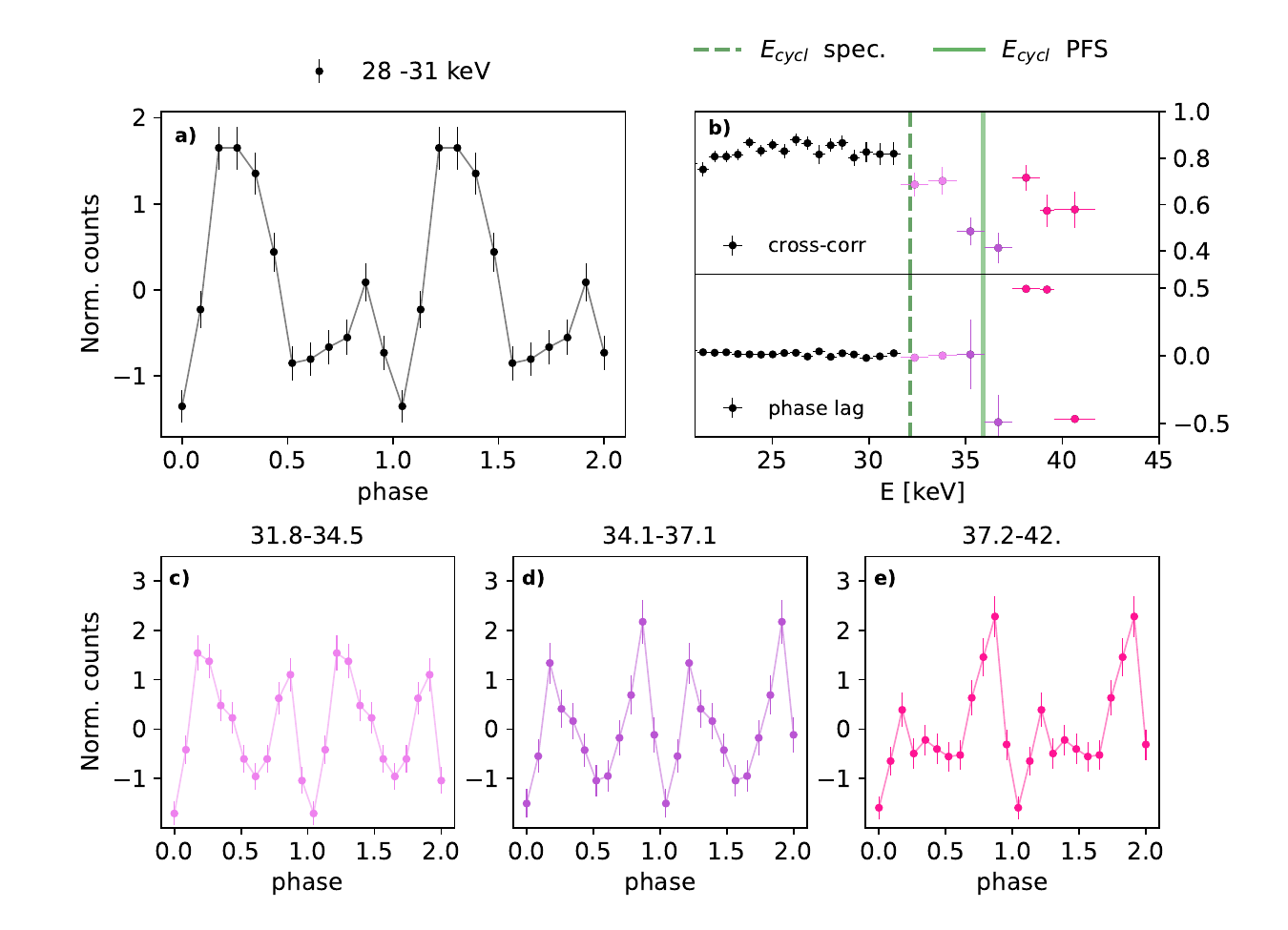}
    \centering
    \includegraphics[height = 0.43\textheight]{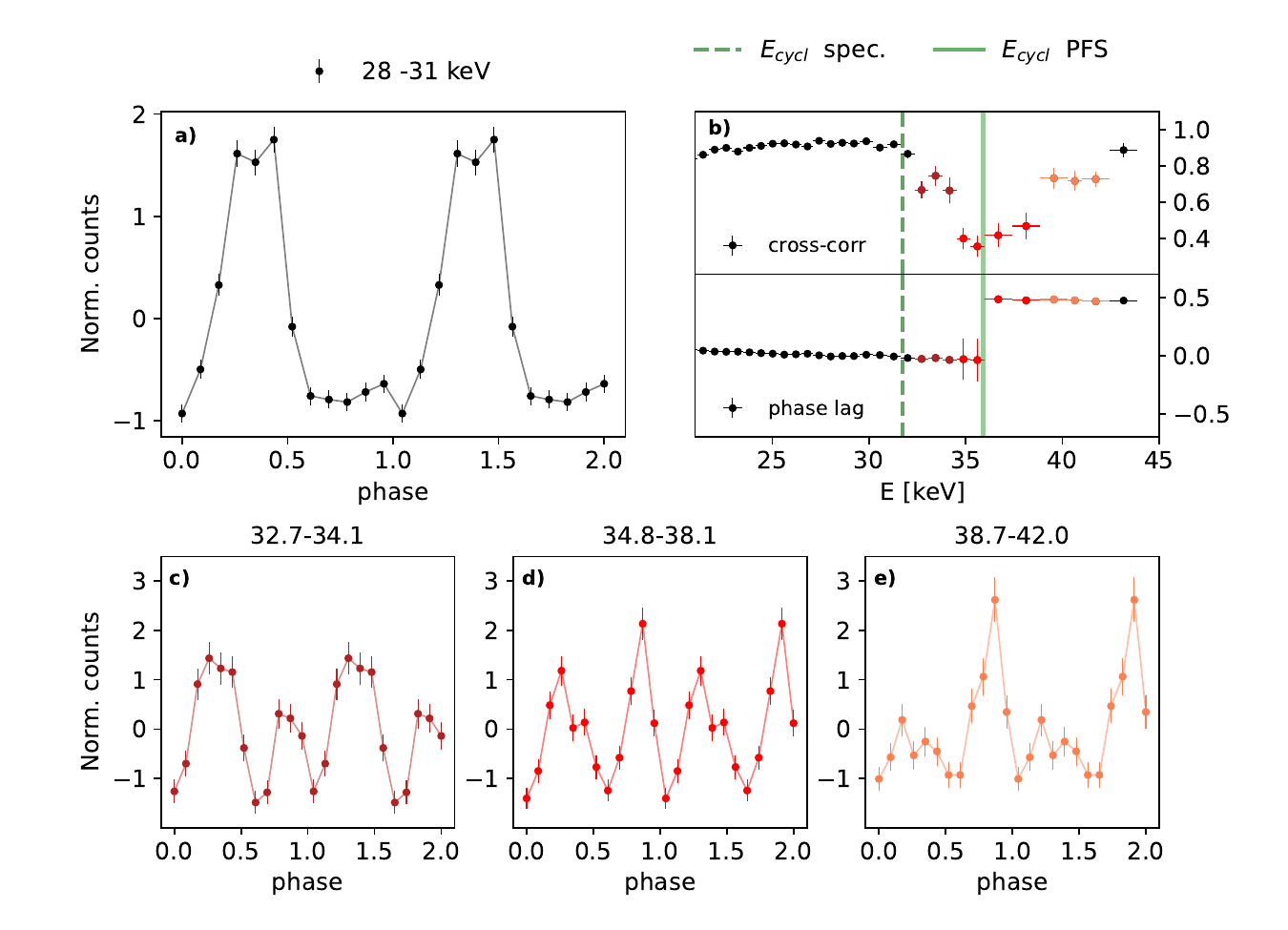}
    \caption{As in Fig.~\ref{fig:pps}, for observations C (top) and D (bottom), including the phase lag spectrum in panel b).}
    \label{fig:cross_a_b}
\end{figure*}
At this stage, a phase-resolved spectral analysis would have been useful to investigate the spectral variability along the spin phase and relate it to the behaviour of the PFS. However, as the signal-to-noise ratio (S/N) is already low in the phase-averaged spectra above 30 keV, we opted to obtain guidance from pulse-profile variability.
The MWCC (Fig.~\ref{fig:epmaps_1}) shows that local profile changes emerge at $\sim28$\,keV; we therefore used the $28$–$31$\,keV profile (\pphighen) as a reference to compute cross-correlation coefficients and phase lags for all profiles above $31$\,keV.
Panels (a) in Figs.~\ref{fig:pps} and \ref{fig:cross_a_b} show \pphighen\ and their cross-correlations (panel b), color-coded by correlation value; panel (b) of Fig.~\ref{fig:cross_a_b} also includes the phase lags. Panels (c)–(e) present average profiles over energy intervals grouped according to the cross-correlation, using the same color coding as panel (b). Dashed and solid green lines mark the centroids of \ecycl\ and $E_{\mathrm{gau,PFS}}$, respectively.
These complementary techniques reveal sharp profile variability between $30$ and $35\,$~keV, particularly prominent at low luminosities. We are led to interpret such abrupt morphological changes, as due to the CRSF, where the energy-dependent cross-section has strong resonances that might reflect in the beam pattern and origin strong variations of the pulse-profile shape.
\section{Discussion}\label{sec:discussion}
In this paper, we analysed for the first time the PFS, and the energy-resolved PPs, along the outburst of one BeXRB, \source\, adopting the same methodology of \cite{ppanda1}. We also applied Bayesian spectral analysis to search for the line in the phase averaged spectra motivated by the PPs and PFS results.\\

\noindent The sharp variability of the PPs around 30 keV (see Fig.~\ref{fig:epmaps_1}) can be interpreted as the direct manifestation of the CRSF: cyclotron resonance introduces a sudden increase in resonant photon-electron scattering cross-section, which depends on the photon propagation angle relative to the magnetic field. This angular dependence redistributes photons at the resonance energy, modifying the observed PPs. And in fact, this relashionship has been confirmed by observations (see, \citealt{ferrigno:2011}, FDA23) and also has been grounded by theoretical and computational studies \citep{schonerr:2014}. In addition, recent developments on modelling have been able to qualitatively reproduce the observed PPs, and the effects of spin phase spectral variability on the PFS at the CRSF energy \citep{ppanda2, ppanda3}.\\

\noindent Focusing on the general trend of the PFS, we note it is closely related to the source flux.  Obs A and B have similar fluxes and very similar PFS energy dependence. The PFS of Obs C, however, displays a bump at a lower energy than the $E_{\text{cycl}}$ derived from spectral analysis, a feature also reported in \cite{ppanda3} for 4U 1538–52. 
Conversely, the PFS of Obs D is similar to those analysed by \cite{ppanda1}, showing a rising trend with energy and a well-defined Gaussian-like dip (see Tab.~\ref{tab:PFSfit} and Fig.~\ref{fig:fit_pfAndharms}). 
Following \cite{ppanda3}, the presence of either a dip or a bump in the PFS depends on the phase-dependent variation of the CRSF. Therefore, the distinct PFS behaviour observed suggests a change in the beam pattern as a function of luminosity. A similar flux-dependent trend of the PFS of \source\ was reported by \citealt{chen08_PFS} for the 2003 outburst; specifically, the middle panel of their Fig. 4 displays a PFS, at a flux comparable to Obs C, with a flat trend in the 10–20 keV range followed by a bump, matching our findings. This comparison supports the interpretation that the observed behaviour is intrinsic to the source's emission mechanism at specific luminosity regimes.\\

\noindent We find evidence for a Gaussian-like absorption feature at
$\approx 32$\,keV in all four observations (Tab.~\ref{tab:bayesian_fit}), with strong to decisive support in three cases (B,C and D) and moderate evidence in the remaining one (A). The centroid energy does not vary significantly, $\approx 1.6\%$, despite a factor of three change in luminosity, while the line depth is luminosity dependent. Our results are partially in contrast with those of \cite{Beri2021}, who detected the line in the phase-averaged spectra only for low-luminosity observations, and only in specific phase-resolved intervals for high-luminosity states. \cite{nabizadeh:2021} reported the line in all phase-averaged spectra; while their results for the two low-luminosity observations agree with ours, they found higher centroid energies (\ecycl\,  $\approx$\,39 and 35 keV) and widths ($\sigma \approx 10$ and 9.9 keV) for Obs A and B, respectively. In their analysis, the centroid energy appears to vary with luminosity. In contrast, in our study \ecycl\ centroid remains consistent within uncertainties across all four observations. Although a luminosity-dependent variation of the line, as observed in other sources, cannot be excluded, the limited statistics above 40 keV prevented us from testing this scenario further.
A shallow dip in the energy spectrum can be observed at the intersection of two continuum components during low-luminosity states ($L_x < 10^{35}$\, \unilum), as previously reported for several sources \citep[e.g.][and references therein]{malacaria:2026}. However, theoretical simulations demonstrate that resonant scattering induces a sharp angular redistribution of photons, forcing them to escape preferentially perpendicular to the magnetic field lines at the CRSF energy \citep[see, e.g.][]{schonerr:2014}. This mechanism results in significant energy-dependent variability in the pulse profiles.
Consistent with these findings, our timing analysis revealed sharp, narrow-range features in both the cross-correlation and Pulsed Fraction (PF) spectra. These features were observed at approximately the same energy, regardless of the varying luminosity levels or changes in the underlying spectral shape.

\noindent A weak luminosity dependence of the cyclotron line energy might be possible when a source is observed close to the critical luminosity, where the accretion flow transitions between sub- and super-critical regimes and the emission geometry changes from a pencil- to a fan-beam configuration \citep{becker:2012,mushtukov:2015}. Such behaviour has been observed in systems like V 0332+53 and A 0535+262, where $E_{\mathrm{cyc}}$ remains approximately constant over a limited luminosity range around the transition \citep{tsygankov:2010,kong21}. However, while the energy remains consistent within uncertainties, we observe an anti-correlation between the line optical depth and luminosity: the line is shallower at the outburst peak and deepens as the luminosity declines. This behaviour can be explained invoking transition from a radiation-pressure-dominated flow, where multiple resonant scatterings in a dense column results on a  "fill in" of the absorption feature \citep{nishimura:2014}, to a lower-luminosity state  which produces a cleaner, more pronounced line profile. A similar conclusion was also reached by \cite{2020JHEAp..27...38T}, based on Insight-HMXT observations taken outside the outburst peak, although through different arguments.

\noindent For an observed cyclotron line energy of $E_{\text{cycl}} \approx 32$~keV, the surface magnetic field is estimated to be $B_{12} \in [3.3, 3.9]$.\footnote{$B_{12} = (1+z) \times E_{\text{cycl}}(\text{keV})/11.57$, assuming a gravitational redshift $z \in [1.2, 1.4]$ for typical neutron star parameters.} This corresponds to a critical luminosity of $L_{\text{crit}} \approx 1.5 \times 10^{37} B_{12}^{16/15} \approx (5.4 - 6.4) \times 10^{37}$~erg\,s$^{-1}$.
Given the lack of consensus on the distance to 4U 1901+03, we adopted a conservative range of $7\text{--}12.5$~kpc to estimate a luminosity range.\footnote{Distance estimates vary significantly: Gaia EDR3 suggests $4.5\text{--}5.6$~kpc, though the parallax is poorly constrained ($\varpi \approx 0.0005 \pm 0.2$~mas). While \cite{2020JHEAp..27...38T} proposed $\approx 12.4$~kpc via torque modelling, \cite{stierhof:2025} argue that current analytic torque models are subject to significant systematics and may not yet allow for reliable quantitative measurements of the magnetic field or associated parameters. We thus treat the torque-derived value only as a tentative upper limit. Conversely, the high source extinction ($E(B-V) \approx 2.5\text{--}3.5$) favors a distance of at least $\approx 7$~kpc, where Galactic dust maps (e.g., Bayestar2019) typically saturate.} Within this interval, the luminosity ranges (for fluxes reported in Tab.~\ref{tab:log}) are : 
$L_{\text{A}} \in [4.1, 13.0]$, 
$L_{\text{B}} \in [4.7, 15.0]$, 
$L_{\text{C}} \in [2.6, 8.4]$, and 
$L_{\text{D}} \in [1.6, 5.1]$ (in $10^{37}$~erg\,s$^{-1}$ units). 
Within a distance of d $\approx 10$~kpc, observation C would be in the range of critical luminosity for the estimated magnetic field.
\section{Conclusions}
\label{Sec:sum_and_concl}
In this paper, we showed how a synergic view of timing and spectral features align to build a coherent picture for the presence of the CRSF in this source.
We have exploited timing and spectral analyses in a complementary manner, showing that the pulsed profiles display a persistent, energy-localized variation that is consistently observed across the four observations through a moving cross-correlation analysis. This behaviour represents a robust, model-independent signature of the imprint of the cyclotron resonant scattering feature on the pulse profiles. By using timing information to identify departures from a simple power-law continuum at the corresponding energies, we suggest that the source is close to the critical luminosity undergoing a transition evidenced by hanges in the width and depth of the cyclotron scattering feature.

\begin{acknowledgements}
E.A. thanks Melania Del Santo, Domitilla De Martino and Valentina La Parola for insightful discussions and thoughtful comments that helped improve this work.
The Authors thank the referee for their very constructive report.
EA acknowledges support from the INAF MINI-GRANTS number 1.05.23.04.04 for the project PPANDA (Pulse Profiles of Accreting Neutron Stars Deeply Analysed). 
This research was supported by the International Space Science Institute (ISSI) in Bern, through the ISSI Working Group project \href{https://collab.issibern.ch/neutron-stars/}{Disentangling Pulse Profiles of (Accreting) Neutron Stars}.

EA, AD, GC acknowledge funding from the Italian Space Agency, contract ASI/INAF n.I/004/11/4. 

AD and FP acknowledge support from SEAWIND grant funded by the European Union - Next Generation EU, Mission 4 Component 1 CUP C53D23001330006.

FP and AD  acknowledge the INAF GO/GTO grant number 1.05.23.05.12 for the project OBIWAN (Observing high B-fIeld Whispers from Accreting Neutron stars).

A.T. acknowledges supercomputing resources and support from ICSC - Centro Nazionale di Ricerca in High Performance Computing, Big Data and Quantum Computing - and hosting entity, funded by European Union - NextGenerationEU.
E.S.L. acknowledges support from the FAU Emerging Talents Initiative and funding from the European Union’s Horizon 2020 programme under the AHEAD2020 project (grant agreement No. 871158).

We made use of Heasoft and NASA archives for the \nustar data.
We developed our own timing code for the epoch folding, orbital correction,
building of time-phase and energy-phase matrices.
This code is based partly on available Python packages such as:
\texttt{Astropy} \citep{astropy:2013, astropy:2018, astropy:2022},
\texttt{lmfit} \citep{lmfit},
\texttt{matplotlib} \citep{Hunter:2007},
\texttt{emcee} \citep{emcee},
\texttt{stingray} \citep{stingray1,stingray2},
\texttt{corner} \citep{corner},
\texttt{scipy} \citep{scipy}.

%     EA acknowledges grants, people ... 
\end{acknowledgements}
\bibliographystyle{aa}
\bibliography{used_only}
\onecolumn    
\appendix

%---------------------------------------------------------------


\end{document}